\documentclass[sigplan,10pt]{acmart}

\usepackage[compact]{titlesec}

\titlespacing\section{0pt}{6pt plus 2pt minus 2pt}{2pt plus 2pt minus 2pt}
\titlespacing\subsection{0pt}{4pt plus 2pt minus 2pt}{2pt plus 1pt minus 1pt}
\titlespacing{\paragraph}{0pt}{3pt plus 0pt minus 1pt}{1.0ex}

\usepackage[subtle]{savetrees}

\usepackage{setspace}

\usepackage{paralist}
\usepackage{float}
\usepackage[utf8]{inputenc}
\usepackage{graphicx}
\graphicspath{ {./include/} }
\usepackage{cleveref}
\usepackage{color}
\usepackage{tcolorbox}
\usepackage{multicol}

\usepackage{float}
\newfloat{lstfloat}{tbp}{lop}
\floatname{lstfloat}{Listing}

\usepackage{listings}
\definecolor{lightgrey}{RGB}{244,244,244}
\definecolor{darkgrey}{RGB}{99,99,99}

\copyrightyear{2023} 
\acmYear{2023} 
\setcopyright{rightsretained} 
\acmConference[PLOS '23]{12th Workshop on Programming Languages and Operating Systems}{October 23, 2023}{Koblenz, Germany}
\acmBooktitle{12th Workshop on Programming Languages and Operating Systems (PLOS '23), October 23, 2023, Koblenz, Germany}\acmDOI{10.1145/3623759.3624550}
\acmISBN{979-8-4007-0404-8/23/10}

\begin{document}

\title{Software Compartmentalization Trade-Offs\\with Hardware Capabilities}

\author{John Alistair Kressel, Hugo Lefeuvre, Pierre Olivier}
\affiliation{%
 \institution{The University of Manchester}
  \city{Manchester}
  \country{UK}
}


\renewcommand{\shortauthors}{Kressel, et al.}


\begin{abstract}
Compartmentalization is a form of defensive software design in which an application is broken down into isolated but communicating components.
Retrofitting compartmentalization into existing applications is often thought to be expensive from the engineering effort and performance overhead points of view.
Still, recent years have seen proposals of compartmentalization methods with promises of low engineering efforts and reduced performance impact.
ARM Morello combines a modern ARM processor with an implementation of Capability Hardware Enhanced RISC Instructions (CHERI) aiming to provide efficient and secure compartmentalization.
Past works exploring CHERI-based compartmentalization were restricted to emulated/FPGA prototypes.

In this paper, we explore possible compartmentalization schemes with CHERI on the Morello chip.
We propose two approaches representing different trade-offs in terms of engineering effort, security, scalability, and performance impact.
We describe and implement these approaches on a prototype OS running bare metal on the Morello chip, compartmentalize two popular applications, and investigate the performance overheads.
Furthermore, we show that compartmentalization can be achieved with an engineering cost that can be quite low if one is willing to trade off on scalability and security, and that performance overheads are similar to other intra-address space isolation mechanisms.
\end{abstract}

\maketitle
\pagestyle{plain}

\section{Introduction}

Software compartmentalization is one of the ways to enforce the principle of least privilege~\cite{Saltzer1975}.
Compartmentalization enforces isolation between components of a software system, granting compartments only the minimal privileges they need to function.
If a component of a compartmentalized system is subverted, the damage the attacker can do is limited to the privileges granted to the compromised compartment~\cite{PAKARGER1987, NPROVOS2003}.
Contrary to many other protection techniques, compartmentalization allows defending against yet unknown/future vulnerabilities in existing code bases~\cite{CHERI_SP}.
Many approaches have been proposed in recent years, utilizing different hardware and software isolation mechanisms to compartmentalize libraries~\cite{codejail, ERIM, Hodor, Narayan2020, Donky, PtrSplit, Cali, CubicleOS, FlexOS, compartos, capvms} as well as smaller pieces of code such as functions~\cite{SOAAP, virtines, intraunikernel, Agadakos2022}.

Morello~\cite{MorelloISA} is an extension to the ARMv8-A architecture implementing the Capability Hardware Enhanced RISC Instructions (CHERI), designed specifically to enable high-performance and scalable compartmentalization~\cite{riscrisk, CHERI_SP, Watson2016}.
This is achieved by enforcing compartment bounds on most memory loads and stores in hardware, and letting communicating compartments securely lend memory to each other using so-called hardware capabilities, a mechanism similar to fat pointers~\cite{lowfatptrs} implemented in hardware to restrict accesses to shared memory at a fine (byte-level) granularity.

When retrofitting compartmentalization to existing code bases, a key challenge is keeping refactoring costs low~\cite{darpa}.
This is crucial not only for reducing the cost of deployment, but also to reduce the number of errors made during the compartmentalization, which can undermine its efficiency or security guarantees~\cite{conffuzz}.
Work exploring compartmentalization with CHERI is so far limited to solutions with relatively high porting costs~\cite{capvms, CHERIJNI, CHERI_SP}, that require a non-negligible reworking to the code corresponding to inter-compartment communications.
These existing works are further limited to MIPS/RISC-V emulated or FPGA prototypes, making it hard to understand the real-world performance one would observe on an ASIC processor.
In that context, the recent availability of Morello raises the research questions we tackle in this paper:
\begin{enumerate}
  \item Which compartment models are possible using Morello, using what programming abstractions, at which refactoring costs?
  \item How does Morello's compartmentalization performance and security guarantees compare to other intra-address space compartmentalization mechanisms (e.g., MPK)?
\end{enumerate}

For this purpose, we adapt an existing compartmentalization-oriented library OS (libOS), FlexOS~\cite{FlexOS}, to Morello, and extend it by developing two compartmentalization programming abstractions relying on hardware capabilities, each representing a particular trade-off in terms of porting costs, security guarantees, and scalability to multiple compartments.
The first is based on manual sandboxing as advocated by CHERI's designers~\cite{CHERI_SP}, with every shared buffer protected by a capability.
Further, we propose a second approach relying on a single region of shared data between two mutually distrusting compartments.
These abstractions are used to compartmentalize popular open source software, SQLite~\cite{SQLITE} and LibSodium~\cite{LIBSODIUM}, at different isolation granularities: functions and libraries.
We evaluate the porting costs, degree of security of these solutions, and further evaluate their performance when executing on the Morello chip, comparing these results to that of another intra-address space isolation mechanism: Intel MPK.
We show that manual porting as advocated by CHERI's designers offers good performance, a good level of security, and scales well to high numbers of compartments.
However, it can require a significant engineering effort when applied to large compartments.
The second approach trades off security guarantees and scalability to more than 2 compartments, to achieve low porting costs, requiring only annotations indicating shared data at declaration time in the code.

\section{CHERI Hardware Capabilities}
\label{sec:bg}

A hardware \textit{capability}~\cite{riscrisk} is an architectural data type used to represent a contiguous region of virtual memory with byte-level granularity.
CHERI hardware capabilities define a base address, bounds and permissions information.
The capability can be dereferenced to access the memory it refers to, with the hardware performing bounds and permissions checks.
Capabilities are made unforgeable by a validity tag, stored separately, and the restriction that capabilities can only be used/manipulated via capability-aware instructions.

When using Morello for compartmentalization (commonly referred to as \emph{hybrid} mode), all compartments share a single address space, and the vast majority of the program's machine code is unchanged, consisting of traditional ARMv8 instructions.
Every memory access made by a core is constrained by two global capabilities that delimit the memory regions the currently executing compartment can access: the \emph{Program Counter Capability} (PCC) and the \emph{Default Data Capability} (DDC).
There is one PCC and one DDC register per core holding these, restricting the ARMv8 code's ability to perform instruction (PCC) and data (DDC) memory accesses within the relevant bounds.
Additional capabilities can be used for sharing data between compartments.
That way, the caller is lending access to the smallest region of memory (the data structure) needed by the callee.
This is secure due to the fine-grained bounds enforcement and efficient as no data copy happens.
Capabilities are also used to control exception-less security domains (compartment) switches, realized by a privileged security monitor (referred to as the \emph{switcher} in this paper).
This is achieved through a special type of capability referred to as \emph{sealed}, which is immutable and non-dereferenceable, and can only be unsealed via a jump to a pre-determined instruction in the switcher.

\section{Design}

We propose two design approaches suitable for compartments on Morello, guided by four main considerations: the engineering effort to retrofit compartmentalization in existing software, compartmentalization's performance overhead, the security of the given approach, and how it scales to many compartments.

\emph{Engineering effort} represents the effort to retrofit isolation into legacy software, or to write new compartmentalized software, using a given abstraction.
It consists of marking compartment boundaries and shared data~\cite{FlexOS}, e.g. with annotations, but also sometimes redesigning part of the software with security in mind~\cite{conffuzz}.
If high, it can be a significant barrier to the adoption of compartmentalization, as it increases costs and development complexity~\cite{FlexOS}.
\emph{Performance overheads} are another factor hindering the popularity of compartmentalization~\cite{ERIM, norton2016hardware} and should be minimized.
For example, the recent DARPA Compartmentalization and Privilege Management call~\cite{darpa} specifies this overhead to be <5\% for application-level compartmentalization with function granularity compartments.
Higher overhead is allowed for proportionally higher security gains.
\emph{Security} is a spectrum of guarantees over an uncompartmentalized system.
The precise security requirements must be judged against the other requirements to strike an acceptable balance.
As a minimum, solutions must enforce strong isolation between compartments and provide access to only a subset of data which has been selectively shared for communication.
Finally, the \emph{scalability} of a solution denotes its capacity to efficiently scale to many compartments.

\subsection{Design Overview}
\label{def:switcher}
In line with Morello/CHERI single address space compartmentalization model~\cite{CHERI_SP}, and with many existing works~\cite{FlexOS, CubicleOS, capvms, intraunikernel, ERIM, Hodor}, we assume a libOS-based environment in which two or more user space and/or kernel components share a single address space and are isolated from each other.
Compartments are defined statically at build time: each compartment is given a pair of memory regions, each contiguous in the virtual address space, to hold its private 1) code and 2) data.
Global compartment capabilities constraining each compartment's memory accesses to the corresponding pair of regions are initialized at boot time.
The static memory layout of the application and the systems software's dynamic memory allocation primitives (\texttt{malloc}/\texttt{mmap}/\texttt{brk}) are designed to allocate data accordingly.
All data declared in the scope of a given compartment is treated as private to that compartment, unless it is specifically annotated as shared.
Shared data is managed differently in the two abstractions we propose, and this is presented in detail in the next subsection.

Gates are inserted in the code in place of function calls, where these calls now cross compartment boundaries.
They invoke the switcher, which performs, on the relevant CPU core, the security domain switch by switching the stack and global compartment code/data capability registers.
Finally, a trampoline is called to enable fast return to the caller compartment from capability-unaware code, with less overhead than re-invoking the switcher.
The switcher is trusted and privileged, hence its code and data (including the global capabilities for all compartments) cannot be accessed by the compartments directly, instead they must use a sealed capability (\S\ref{sec:bg}).
The switching mechanism is kept as lightweight as possible to minimize overhead while preserving strong security.
We define the trusted computing base as the switcher, gates, trampoline, early boot code (including capability initialization code), memory manager, scheduler and interrupt handler.

\subsection{Two Approaches To Sharing Data}

Compartments cannot access memory outside the regions constrained by their \emph{DDC}/\emph{PCC}. This raises the issue of how to selectively share data between compartments, and how to do so efficiently (i.e. without data copy).
We discuss two approaches to data sharing, and reason about their performance, security, and scalability properties.

\subsubsection{Approach 1: Replacing Pointers with Capabilities.}
With CHERI, certain pointers can be transformed into fine-grained capabilities encompassing only the pointed data structure(s).
This is the standard way to manage shared data as designed by the CHERI/Morello authors~\cite{CHERI_SP}.
A compartment \emph{C1} wishing to share a subset of its dedicated memory region with another compartment \emph{C2} can pass such a fine-grained capability \emph{cap} as parameter/return value of a gate.
Morello is designed so that the memory access made by \emph{C2} dereferencing \emph{cap} will not be subject to \emph{C2}'s global capability: \emph{C2} will thus be able to access \emph{C1}'s memory region, but only the few bytes represented by \emph{cap}.

\begin{lstfloat}
\begin{lstlisting}[language = C]
void foo(mystruct *stat, int index) { // Original
  char *str = stat->str;
  bar(str);
  float *element = stat->array[index];
  float *next = element+1;
}

void foo(mystruct *__capability stat, int index) { // Ported
  char *__capability str = stat->str;
  bar((__cheri_fromcap char*)str);
  float *__capability element = stat->array[index];
  float *__capability next = stat->array[index+1];
}
\end{lstlisting}
  \vspace{-.2cm}
  \caption{Example of a function annotated to use capabilities.}
  \label{lst:cap_annotation}
\end{lstfloat}
 
\paragraph{Engineering Cost.}
The sandboxing effort of a legacy function \texttt{foo} is illustrated on Listing~\ref{lst:cap_annotation}.
As described earlier, pointer parameters (\texttt{stat} line 8) and any pointers created out of a capability (\texttt{str} line 9, 11, 12) must be transformed into capabilities with annotations.
Capabilities flowing out of the compartment must be changed back into pointers with a cast (line 10).
Capability monotonicity must also be respected, i.e. a capability cannot extend the bounds of the capability it is derived from: \texttt{element+1} (line 5) is forbidden because it refers to memory outside the bounds of \texttt{next}, and that code must be redesigned.
Other types of changes may be needed depending on the ported code~\cite{CHERI_PROGRAMMING_GUIDE}.
In the general case, we hint that such manual porting may only be amenable to small-scale scenarios (e.g. sandboxing one or a few functions), because the engineering cost of rewriting pointers into capabilities becomes too high as compartments' sizes increase.

\paragraph{Trust Model.}

Engineering costs can be kept low with this approach in scenarios with 1) small compartment sizes, which limits the amount of code rewriting within the compartments; and 2) no capability flowing outside the compartment, to avoid costly rewriting of the data flow in the rest of the application.
This fits very well function sandboxing scenarios.
The isolated function represents a distrusted compartment, and is isolated from the rest of the system.
Pointer arguments entering the compartment are replaced with capabilities.
The rest of the system is trusted and can access any memory within the sandbox compartment.

\paragraph{Performance, Security, and Scalability.}
Data is shared through capabilities, leading to a low performance impact (no copy or marshaling).
In terms of security, the sandbox is isolated by the \textit{PCC}/\textit{DDC}, constraining all non-capability operations made by the sandbox, and shared data is tightly bounded by argument capabilities, resulting in strong isolation.
Regarding scalability, this approach scales to an unlimited number of compartments (e.g. function sandboxes) with a constant porting effort (porting complexity does not grow with the number of compartments).

\subsubsection{Approach 2: Overlapping Shared Region.}
\begin{figure}
\centering
\includegraphics[width=0.4\textwidth]{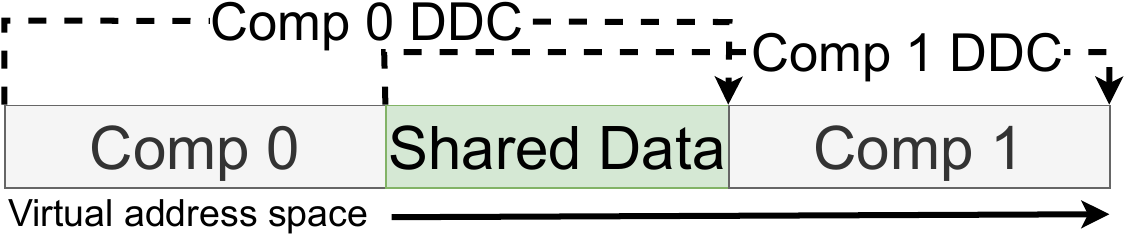}
\caption{Compartment bounds with shared memory regions. The compartment bounds overlap to encompass shared data.}
\label{fig:2comps}
\end{figure}

This approach drops the fine-grained capabilities to rely on a single region of shared data.
The \textit{DDC} bounds of communicating compartments are extended to cover this region, so both can access shared data, as illustrated in \Cref{fig:2comps}.
Since capability bounds must cover contiguous memory, the shared data region is located between two compartments in memory.
The linker and dynamic memory allocation primitives ensure that shared data is correctly placed in the relevant memory.

\paragraph{Engineering Cost.}
With this approach shared data needs to be marked as such with annotations in the source code.
The function calls at compartment boundaries are similarly annotated.
This is illustrated on Listing~\ref{lst:shared_data}, where \texttt{foo} and \texttt{bar} are placed in different compartments.
Code transformations use these annotations to automatically allocate shared data in memory which is accessible from all compartments, and to instantiate gates.
The engineering effort of this approach is relatively low, and significantly lower than replacing pointers with capabilities, as data must only be annotated at declaration and/or allocation sites.

\begin{lstfloat}
\begin{multicols}{2}
  
\begin{lstlisting}[language = C]
void foo() { // Original
  int x;
  bar(&x);
}

void foo() { // Ported
  int __shared x;
  __gate(bar, &x,
    compartment1);
}
\end{lstlisting}
\end{multicols}
\vspace{-.6cm}
\caption{Example of annotations for shared regions.}
\label{lst:shared_data}
\end{lstfloat}

\paragraph{Trust Model.}
Mutual distrust is enforced between compartments, with none able to access the others' private data.

\paragraph{Performance, Security, and Scalability.}
Unlike Approach 1, shared stack variables must be allocated on a heap in the shared region, resulting in an additional allocation cost.
Techniques to address this problem like data shadow stacks~\cite{FlexOS} cannot be applied as-is due to the requirements of the \emph{DDC}.
Nevertheless, we expect performance to be comparable to the previous approach.
Regarding security, isolation of memory accesses is also enforced by the compartment \textit{PCC} and \textit{DDC}.
Data sharing is however made at a coarser granularity, with the entire shared memory region accessible to both compartments at all times. This trades off security in two ways: 1) bounds are not tight to individual objects, thus not offering CHERI's spatial safety for shared objects; 2) even assuming no revocation in Approach 1, the number of objects effectively accessible by each compartment at any execution time will remain larger, resulting in more potential for compartment interface vulnerabilities~\cite{conffuzz}.
In terms of scalability, this approach only scales to a small number of compartments: indeed one can create only a single overlapping region per pair of communicating compartment, hence a scenario with e.g. 3 compartments wishing to access a shared data structure is not possible.

\section{Implementation}
We have selected FlexOS~\cite{FlexOS, FLEXOS_HOTOS}, a compartmentalization-focused library operating system, to implement a prototype system.
FlexOS originally supported isolation with Intel Memory Protection Keys and Extended Page Tables.
The OS allows easy extension to new isolation mechanisms.
Further, its design is based on the Unikraft~\cite{UNIKRAFT} unikernel, so it inherits its high performance, small attack surface, and good compatibility with popular applications.
We ported FlexOS to the Morello platform, and implemented on top of it the two compartmentalization abstractions described earlier, in a total of about 2200 lines of code.
Below we give implementation details regarding the system's initialization, the compartments' structure, and the security domain switching process.

\subsection{Compartment Structure}
Compartments are defined at build time in a configuration file provided to the FlexOS build tool.
At link time, isolated data are placed into their respective, separate, non overlapping ELF section with the help of a custom linker script automatically generated by the toolchain.
Non-isolated data are placed into a default compartment.
The linker script also reserves space for dynamically-allocated data (stack and heap) in each compartment's memory.

The compartment switcher's code and data is isolated from all other compartments' code.
This is done to control access to the switcher, which is a privileged entity.
In addition, compartment capability pairs (one \textit{DDC} and \textit{PCC} pair per compartment) are stored in memory which is not accessible from any compartment but that of the switcher.
This is done to avoid a compartment arbitrarily granting itself access to another compartment's memory.

\subsection{Initialization}
Based on the compartment boundaries defined in the linker script, compartments are initialized at boot time: we trust the boot code of FlexOS to initialize compartment capabilities correctly.
Compartment's \textit{PCC} and \textit{DDC} bounds are initialized to cover the statically defined compartment memory region.
Capability bounds can only cover a contiguous region of memory, meaning that all of the code and data of a compartment must be present in contiguous memory.
Once compartment capabilities have been created, they are stored in the memory reserved for compartment capability pairs.

During boot time, a capability pair for the switcher is also initialized.
This pair grants access to the switcher code, and the compartment capability pairs.
To prevent unauthorized execution of the switcher, the capability pair granting access to the switcher is also placed in memory which is out of bounds of any compartment.
To access this pair, a sealed capability is created for each compartment, which is unsealed using a \texttt{lpb} (load pair and branch) instruction.
The sealed capability is thus the only way for compartments to invoke the switcher.
Each compartment is given one such sealed capability.
Finally, each compartment receives a private allocator, which manages the per-compartment portion of the virtual address space previously reserved in the linker script.
Using this allocator, a private stack and heap for each compartment are initialized.

At the end of the boot process, the capability pair for the default compartment is loaded and execution then enters the default compartment.

\subsection{Switching Security Domains}

\begin{figure}
    \centering
    \includegraphics[width=0.35\textwidth]{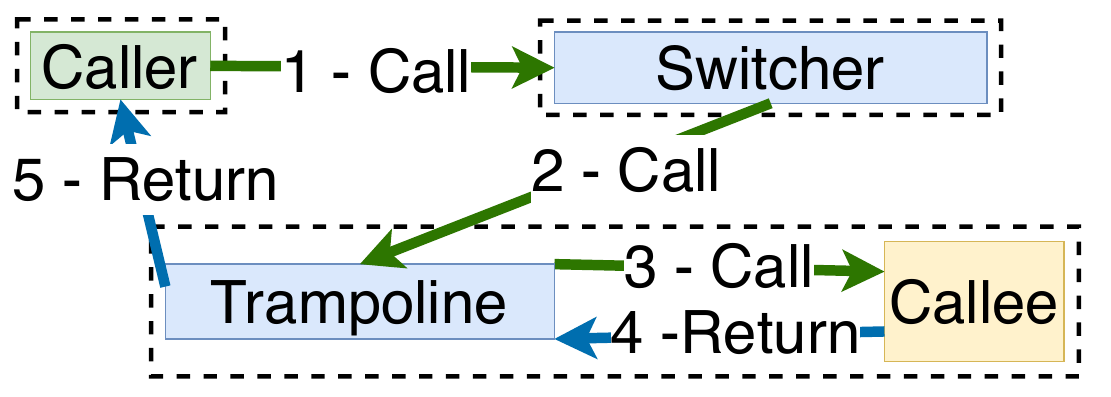}
    \caption{Control flow of a compartment switch (call and return paths).
    Dashed boxes represent protection domains.
    The trampoline is available in the domain of the callee compartment.}
    \label{fig:switch_flow}
\end{figure}

The security domain switch process is illustrated on \Cref{fig:switch_flow}: the caller compartment invokes the privileged switcher safely through a sealed capability, which switches the architectural state representing compartment permissions on the relevant CPU core.
Using a trampoline, the switcher then branches to the callee compartment.
On the return path, a trampoline is used to branch back to the caller.

We implement compartment switch gates as C macros.
This allows the instructions invoking the switcher to be directly inlined at the call site, avoiding the need for a function call.
The call gate is in the caller compartment.
Unlike in other implementations of FlexOS call gates, such as MPK~\cite{FlexOS}, the domain transition is not realized within the security context of the caller: this is done to prevent compartments from accessing the capability pairs of other compartments.
Instead, it invokes the switcher after having loaded the parameters needed for the callee.
When initiating a switch, the compartment switch gate takes the caller and callee compartment IDs, the callee function pointer, a return variable pointer (if needed) and arguments to be passed.

The compartment switch gates follow the AArch64 calling convention for argument registers.
The procedure used to invoke the compartment switcher is as follows: caller-saved registers are pushed to the stack, the current stack and frame pointers are saved, the switcher parameters are loaded and finally, the sealed capability granting access to the switcher capabilities is loaded, unsealed and the switcher is invoked.
By invoking the switcher, the PCC is restricted to only execute switcher code.

The switcher is an isolated entity which is trusted to perform the compartment switches.
The switcher \textit{PCC} is the only capability able to execute switcher code, which is isolated from all other code.
The switcher \textit{DDC} is the only way to access compartment capability pairs.
Once the switcher is invoked, the following steps are taken:

\begin{compactenum}
    \item Upon first entering the switcher, the caller compartment \textit{DDC} is still in place.
    This, along with the return capability generated by the call to the switcher are stored on the caller compartment stack.
    A sealed capability is generated which grants access to this stored capability pair.
    \item The \textit{DDC} is changed to the switcher \textit{DDC}.
    \item Callee compartment capabilities (PCC, DDC) are loaded/set.
    \item The stack is switched for the callee.
    \item The callee compartment \textit{PCC} is used to leave the switcher and jump to the trampoline.
\end{compactenum}

The trampoline serves as both the entry and exit point for a compartment.
Return to the caller compartment can only be performed via the capability pair stored on the caller stack, accessed via a sealed capability. This avoids the need to go through the switcher on the return path.
At call time, the trampoline stores the sealed capability created by the switcher onto the callee's stack before calling the target function.
Upon return to the trampoline, the sealed capability is popped, unsealed, and a return to the caller compartment is performed. The unsealed capability is used to load the caller compartment capability pair: the \textit{PCC} is set as part of the return and the \textit{DDC} is restored by the call gate in the caller. Upon return from the callee compartment, the gate restores the stack and any saved registers.
If the function call returned a value, the gate will store the returned value in a provided variable pointer.

\section{Evaluation}
In this section we evaluate the impact on performance and engineering effort of our proposed approaches.
We use the abstractions we developed to compartmentalize two popular applications, the SQLite~\cite{SQLITE} database management system and the libsodium~\cite{LIBSODIUM} crypto library.
For libsodium we use our first approach to data sharing, function sandboxing with fine-grained capabilities, and sandbox 5 functions manipulating external input, listed in \Cref{tab:portingeffort}.
The library is integrated with a benchmark we derived from its test suite, running representative tests (e.g. encrypting a buffer, generating a key) 200 times in a loop.
SQLite's compartmentalization uses our second approach, coarse-grain shared data regions with overlapping DDCs between communicating compartments.
We create two mutually distrusting compartments: the filesystem management code, and the rest of the system.
This application is benchmarked with 5000 \texttt{INSERT} operations on an in memory (\texttt{ramfs}) database.
The compartmentalization scenario and the benchmark are both taken from the FlexOS paper~\cite{FlexOS} and represent a system call intensive application.

We run all experiments on our port of FlexOS, bare-metal on the Morello evaluation board~\cite{MorelloSDP, MORELLO_BOARD2} with 16 GB of RAM and the capability-enabled SoC clocked at 2.5 GHz.
For comparison, we also gather data for Linux on Morello, running a capability-unaware AArch64 Debian 11, as well as for the other isolation mechanisms supported by FlexOS (MPK, EPT) on an x86-64 Xeon Silver 4114 clocked at 2.2 Ghz with 128 GB of RAM, running Debian 11.
Results are averages of 10 runs; since they show little variance we omit error bars.

\subsection{Engineering Cost}

Table \ref{tab:portingeffort} shows the porting effort associated with the compartmentalization of libsodium and SQLite.

\begin{table}
    \center
    \caption{Porting effort required to compartmentalize.}
    \label{tab:portingeffort}
    \includegraphics[width=.45\textwidth]{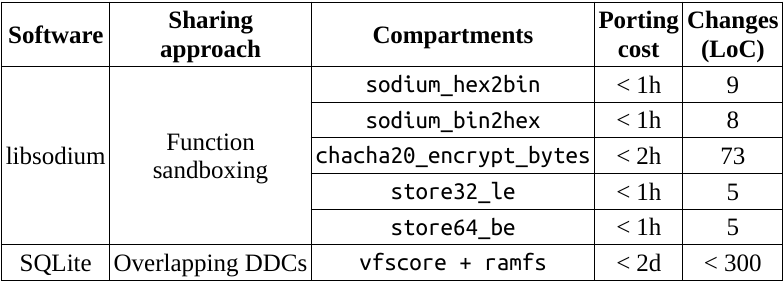}
\end{table}

Concerning libsodium, the engineering cost of sandboxing functions with our first approach involves rewriting the function's code to be capability-aware, something that can be a task of non-negligible complexity~\cite{CHERI_PROGRAMMING_GUIDE}.
Still, because we deliberately selected functions with relatively small sizes (11 to 141 LoC), that effort was relatively low (1 or 2 hours per function for a programmer with good knowledge of capability programming), and mostly consisted of annotating the relevant pointers to be transformed into capabilities.

Regarding SQLite, we achieved the compartmentalization in a couple of days using the overlapping shared data region approach.
The effort involved can be broken down into two tasks: 1) gate insertion and 2) shared data identification.
Gate insertion is mostly automated by the FlexOS toolchain, with the programmer only needing to insert annotations at the desired compartment boundary.
The majority of the work comes from identifying shared data.
It is currently with FlexOS a manual process, during which the programmer must analyze the code carefully to pinpoint what needs to be shared with annotations.
Although the engineering cost for this approach seems higher than for the sandboxing method, the compartment size is also much larger for the overlapping DDC, e.g. 5.8K LoC for the filesystem compartment.

\begin{center}
    \begin{table*}[tbh]
    \centering
        \caption{Performance counters by for each configuration compared to the uncompartmentalized baseline.}
        \label{tab:perf}
    \begin{tabular}{|p{0.22\textwidth} | p{0.06\textwidth} p{0.09\textwidth} p{0.075\textwidth} p{0.09\textwidth} p{0.06\textwidth} p{0.09\textwidth} p{0.075\textwidth} p{0.06\textwidth}|}
     \hline
     Configuration & L1I Acc & L1I Misses & L1D Acc & L1D Misses & Br Ret & Br Mispred & Mem Acc & Inst Ret\\
     \hline
     \texttt{sodium\_hex2bin} \& \texttt{sodium\_bin2hex} & +0.15\% & +3.95\% & +0.09\% & +25.11\% & +0.29\% & +10.25\% & +0.13\% & +1.74\% \\
     \hline
     \texttt{chacha20\_encrypt\_bytes} & +23.58\% & \textbf{+209.49\%} & +16.85\% & \textbf{+106.00\%} & +23.33\% & +76.76\% & +16.69\% & +4.90\% \\
     \hline
      \texttt{store32\_le} \& \texttt{store64\_be}   & +16.28\% & -37.47\% & +13.98\% & +12.03\% & +14.29\% & \textbf{+91.72\%} & +13.84\% & +2.90\% \\
      \hline
      \texttt{libsodium all} & +16.36\% & \textbf{+178.92\%} & +12.44\% & \textbf{+189.79\%} & +15.73\% & \textbf{+102.34\%} & +12.15\% & +4.54\% \\
     \hline
     \texttt{SQLite} & \textbf{+99.6\%} & +46.0\% & +48.1\% & +18.4\% & +6.5\% & \textbf{+175.2\%} & +48.3\% & +27.1\% \\
     \hline
    \end{tabular}
    \end{table*}
\end{center}

\begin{figure}
    \centering
    \includegraphics[width=0.45\textwidth]{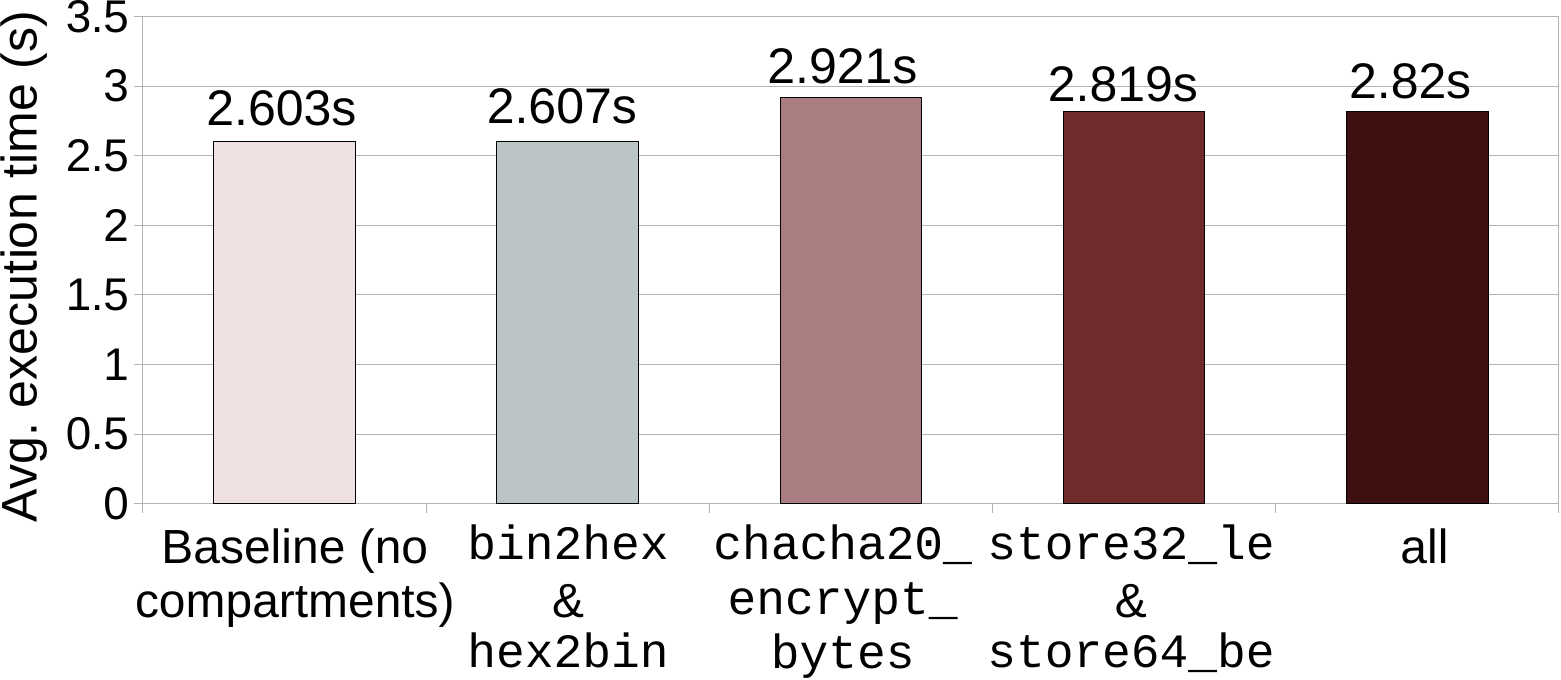}
    \caption{Overhead of various compartmentalization scenario on libsodium (X labels are sandboxed functions).}
    \label{fig:libsodexectime}
\end{figure}

\vspace{-.6cm}
\subsection{Performance} 

\paragraph{Libsodium.}
We analyze different configurations of the 5 \texttt{Libsodium} functions we sandboxed by replacing pointers with capabilities.
The results are presented on \Cref{fig:libsodexectime}.
The overhead is very modest, due to a relatively low amount of compartment switches; the highest is \\ \texttt{chacha20\_encrypt\_bytes} with 0.669 compartment switches/1k instructions.
The lowest performance overhead is achieved when \texttt{sodium\_hex2bin} and \texttt{sodium\_bin2hex} are isolated, adding only a 0.144\% performance overhead.
In contrast, the highest performance overhead comes from compartmentalizing \texttt{chacha20\_encrypt\_bytes} only, with an overhead of 12.207\%.
This is higher than the scenario where all are isolated, because \texttt{chacha20\_encrypt\_bytes} makes calls to \texttt{store32\_le}.
When only \texttt{chacha20\_encrypt\_bytes} is isolated, a compartment switch is required for each call, hence the overhead is higher.
Evaluating against the DARPA requirements for function granularity isolation~\cite{darpa}, most of these results are in range of the required 5\% overhead.

To further understand this behavior, we enabled hardware performance counters support in our prototype OS and gathered the data presented in Table \ref{tab:perf}.
The number of instructions executed, memory accesses performed including cache accesses/misses, and branches executed, increase proportionally.
The reduction in instruction cache misses for \texttt{store32\_le} and \texttt{store64\_be} is due to the new wrapper function being no longer inlined as the original functions were, resulting in more efficient cache utilization.
We also note that the branch predictor struggles with increased use of indirect branches, a direct consequence of gates as a layer of indirection.

\begin{figure}
    \centering
    \includegraphics[width=0.45\textwidth]{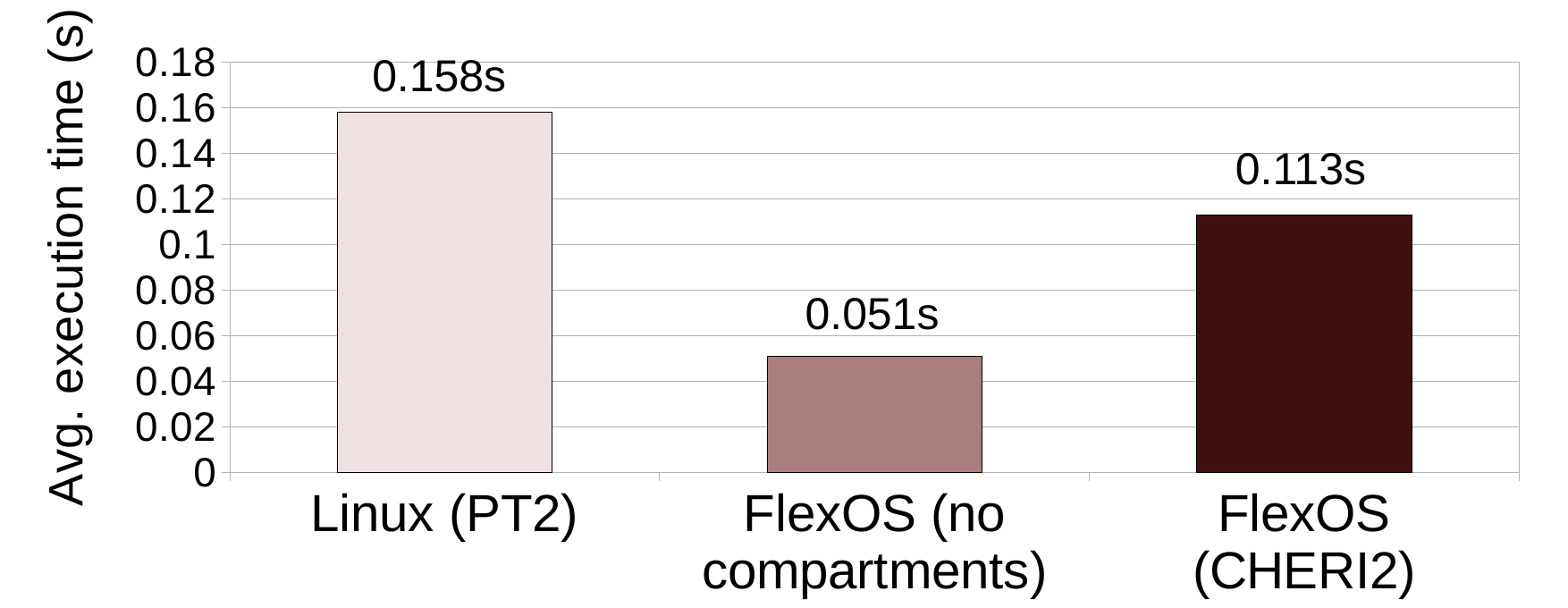}
    \caption{Execution times of different configurations of SQLite running on Morello.}
    \label{fig:sqliteexectime}
\end{figure}

\paragraph{SQLite.}
Results for SQLite are presented on \Cref{fig:sqliteexectime}. Here the overlapping shared data region approach is used for CHERI configurations.
Compared to an uncompartmentalized FlexOS baseline, isolating the filesystem adds an overhead of 119.9\%.
This is because the isolated code lies on the hot path, meaning that isolated primitives are frequently called: we measured a high frequency of domain transitions (2.49/1k instructions).
However, CHERI-based filesystem isolation still outperforms the same benchmark running on an unmodified Debian Linux installation by a factor of 1.4x.
Running on Debian Linux is equivalent to a relatively costly two-compartments page-table-based scenario (PT2), due to the page table-based user-kernel separation.

The data measured from the hardware performance counters for SQLite is presented in Table \ref{tab:perf}.
Compartmentalization increases the number of instructions executed by 27.1\% and memory accesses by 48.3\%.
Correspondingly, the number of L1 instruction cache and L1 data cache access increase, while the number of misses for both increase by a smaller proportion.
Interestingly, the branch misprediction rate rises by a far greater amount than the number of branches executed.
This may be attributed to the increased number of indirect branches used as a result of the switching process, which are harder for the predictor to predict.

\begin{figure}
    \centering
    \includegraphics[width=0.45\textwidth]{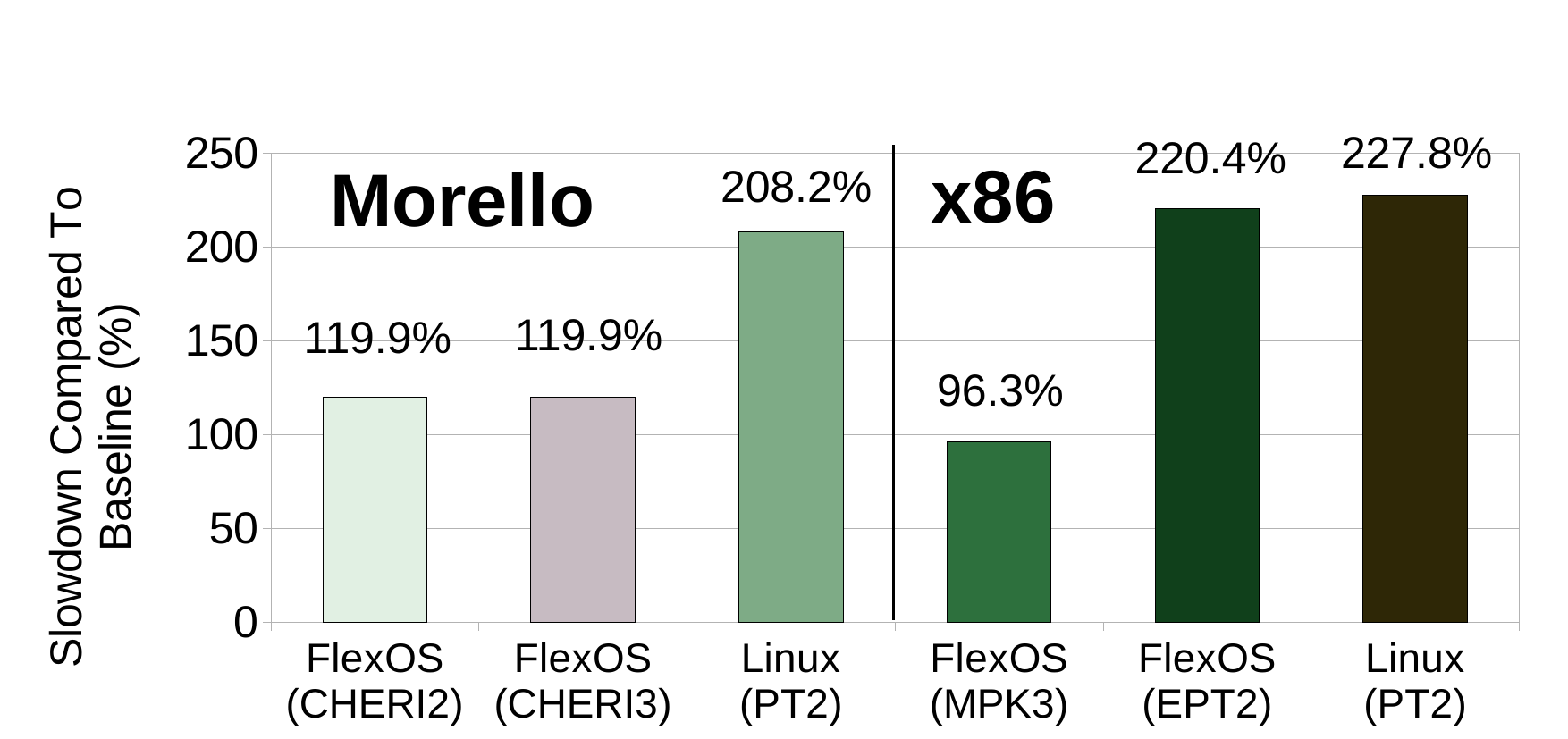}
    \caption{Overhead relative to uncompartmentalized FlexOS for SQLite.}
    \label{fig:sqliterelativeoverhead}
\end{figure}

The relative overhead figures for SQLite are presented in \Cref{fig:sqliterelativeoverhead}.
We also include numbers for FlexOS running on x86-64 with MPK and EPT.
We present results for the previously-mentioned compartmentalization scenario, 2 compartments (filesystem and rest of the system, CHERI2/PT2/EPT2), and an additional scenario with 3 compartments (filesystem, time manager, rest of the system, CHERI3/MPK3).
Morello's numbers are in line with the relative overheads of these other isolation mechanisms.
Compared to the overhead of MPK3, CHERI is slightly more expensive.
This can be attributed to the switching mechanism that with CHERI requires additional bookkeeping and jumps (e.g. to the switcher) compared to MPK.

\begin{figure}
    \centering
    \includegraphics[width=0.5\textwidth]{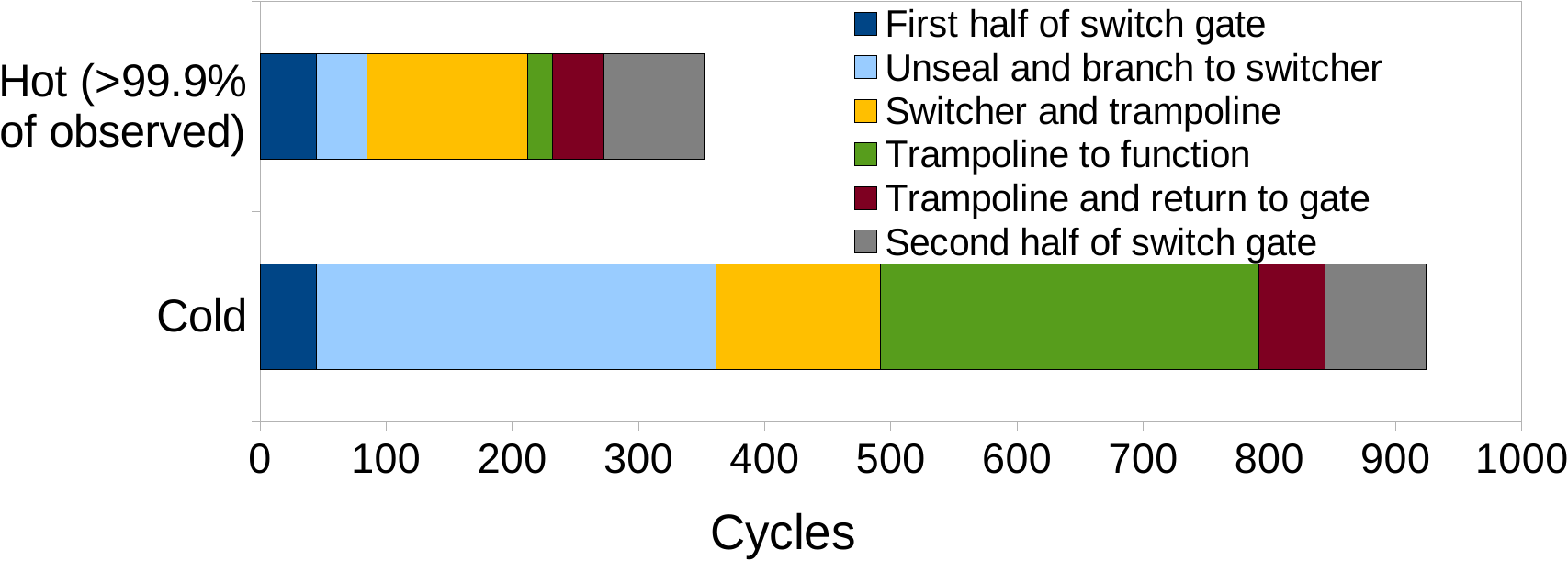}
    \caption{Hot and cold compartment switch latencies, broken down into component parts.}
    \label{fig:microhotcold}
\end{figure}

\paragraph{Security Domain Switch Latency Breakdown.}
We observe that the majority of the overhead comes from security domain switches~\cite{flexsc}, hence we use microbenchmarks to obtain a breakdown of the cost in CPU cycles associated with a switch.
Results are presented in Figure \ref{fig:microhotcold}.
Compartment switches can be broken down into hot and cold switches.
This is a result of cache utilization.
The vast majority of switches (>99.9\%) observed in all configurations fall into the category of hot switches.
The cold switches, therefore, represent a worst-case cycle latency for compartment switches.
These can be expected in compartmentalization scenarios where compartment switches rarely occur, and cache utilization is worse.
This results in a best switching latency of <400 cycles, and a worst case of 900-1000 cycles.

\section{Related Work}

\paragraph{Compartmentalization.} In recent years, many works have looked at implementing various forms of compartmentalization~\cite{Cali, Hodor, PtrSplit, Narayan2020, CubicleOS, capvms, Donky, ERIM, cherios, compartos, FlexOS, wedge, Shreds, intraunikernel, codejail, virtines, SOAAP, hakc, ksplit, privman, privtrans, datashield, nestedkernel}.
Many of these approaches have focused on library isolation~\cite{codejail, ERIM, Narayan2020, Hodor, Donky, PtrSplit, Cali, CubicleOS, FlexOS, compartos, capvms}, while others approach isolation in a much more fine-grained way, including function level isolation~\cite{SOAAP, cherios, intraunikernel, virtines}.
Isolation in single-address-space OSes such as Library OSes has also been explored~\cite{FlexOS, intraunikernel, CubicleOS}, although using other mechanisms such as memory protection keys.

\paragraph{CHERI.} While compartmentalization using CHERI hardware capabilities has been explored in the past~\cite{cherios, compartos, cherirtos}, little work has been done to explore compartmentalization on ASICs, instead using FPGA prototypes.
Cap-VMs~\cite{capvms} provides capability aware VM-like abstractions which can be used to isolate capability-unaware components of a system using CHERI RISC-V.
CHERI JNI~\cite{CHERIJNI} uses capabilities to provide memory safety for the Java VM.
Finally, CheriBSD~\cite{CHERI_SP} uses capabilities to enable (manual) application compartmentalization.

\section{Conclusion}
We have proposed two solutions to implement compartmentalization on top of a real-world hardware capability-enabled SoC, ARM Morello.
Each solution represents a specific point in the compartmentalization design space, and we have explored the trade-offs they represent in terms of engineering effort for retrofitting legacy software, performance overheads, security, and scalability.
We show that hardware capability-based compartmentalization can be achieved with a variable engineering cost, which can be quite low if one is willing to trade-off on scalability and security, and that performance overheads are similar to other intra-address space isolation mechanisms (e.g. memory protection keys), and lower than more heavyweight (page table/extended page table) solutions.
\section{Acknowledgments}
We thank the anonymous reviewers for their insights. This work was partly funded by the EPSRC/Innovate UK grant EP/X015610/1 (FlexCap), the UK’s EPSRC grants EP/V012134/1 (UniFaaS), EP/V000225/1 (SCorCH), a studentship from NEC Labs Europe and a Microsoft Research PhD Fellowship.

\bibliographystyle{plain}
\bibliography{bibliography}

\end{document}